\documentclass[conference]{IEEEtran}
\IEEEoverridecommandlockouts
\usepackage{cite}
\usepackage{amsmath,amssymb,amsfonts,amsthm}
\usepackage{algorithmic}
\usepackage{graphicx}
\usepackage{textcomp}
\usepackage{xcolor}
\usepackage[hypcap=true]{caption}
\usepackage{etoolbox}
\usepackage{euscript}
\usepackage[T1]{fontenc}
\usepackage[utf8]{inputenc}
\usepackage{nimbusserif}
\usepackage{ifpdf}
\usepackage[english]{babel}
\usepackage{subcaption}
\usepackage{color}
\usepackage{graphicx}
\usepackage{hyperref}

\newtheorem{definition}{Definition}
\newtheorem{proposition}{Proposition}
\newtheorem{lemma}{Lemma}

\definecolor{darkviolet}{rgb}{0.58,0,0.83}

\def\BibTeX{{\rm B\kern-.05em{\sc i\kern-.025em b}\kern-.08em
    T\kern-.1667em\lower.7ex\hbox{E}\kern-.125emX}}
\begin{document}

\title{Phase-Based Signal Representations for Scattering\\
\thanks{This work was supported by the Austrian Science Fund (FWF) projects Y 551-N13 (FLAME) and I 3067-N30 (MERLIN).\\
\emph{This is the Author's Accepted Manuscript version of work presented at EUSIPCO21. It is licensed under the terms of the \href{https://creativecommons.org/licenses/by/4.0/}{Creative Commons Attribution 4.0 International License}, which
permits unrestricted use, distribution, and reproduction in any medium, provided the original author and source are credited. The published version is available at:} \url{https://ieeexplore.ieee.org/abstract/document/9616285}}
}

\author{\IEEEauthorblockN{1\textsuperscript{st} Daniel Haider}
\IEEEauthorblockA{\textit{Acoustics Research Institute} \\
\textit{Austrian Academy of Sciences}\\
Vienna, Austria \\
daniel.haider@oeaw.ac.at}
\and
\IEEEauthorblockN{2\textsuperscript{nd} Peter Balazs}
\IEEEauthorblockA{\textit{Acoustics Research Institute} \\
\textit{Austrian Academy of Sciences}\\
Vienna, Austria \\
peter.balazs@oeaw.ac.at}
\and
\IEEEauthorblockN{3\textsuperscript{rd} Nicki Holighaus}
\IEEEauthorblockA{\textit{Acoustics Research Institute} \\
\textit{Austrian Academy of Sciences}\\
Vienna, Austria \\
nicki.holighaus@oeaw.ac.at}
}

\maketitle

\begin{abstract}
The scattering transform is a non-linear signal representation method based on cascaded wavelet transform magnitudes. 
In this paper we introduce phase scattering, a novel approach where we use phase derivatives in a scattering procedure. 
We first revisit phase-related concepts for representing time-frequency information of audio signals, in particular, the partial derivatives of the phase in the time-frequency domain. 
By putting analytical and numerical results in a new light, we set the basis to extend the phase-based representations to higher orders by means of a scattering transform, which leads to well localized signal representations of large-scale structures. 
All the ideas are introduced in a general way and then applied using the STFT. 

\end{abstract}

\section{Introduction}\label{sec:intro}
For most applications using time-frequency representations, only the magnitude of the obtained complex coefficients is considered
as, e.g., in the cases of spectrogram and scalogram \cite{flandrin}. This is a standard method for analyzing the time-frequency content of an audio signal, which is also heavily used in deep learning for audio,
where neural networks and time-frequency (magnitude) representations often go hand-in-hand \cite{Marafiotia,Huang2019b,8678825}. In this context, the \emph{scattering transform} provides an important link, 
as it may be interpreted as a deterministic neural network, computing a cascade of filter magnitude transformations, i.e. a sequence of filtering and application of the modulus as non-linearity. As signal representation, it is able to reveal large-scale structures of the signal \cite{axioms8040106}, which is partly caused by the non-linearity, i.e. discarding the phase information. The phase is visually very unintuitive but its derivative in the time, as well as in the frequency direction yield very informative and intuitive phase-based signal representations. The resulting quantities overcome some of the issues of the (pure) phase and are directly linked to the concepts of \emph{instantaneous frequency} and \emph{group delay}. The interest of this work is to investigate the idea of scattering with a complementary approach as before, namely considering the partial derivatives of the phase instead of the modulus as non-linear step. As in the magnitude case, larger-scale structures can be revealed with this \textit{phase scattering} procedure, additionally inheriting the precision of the localization of time-frequency information that is provided by these phase-based signal representations.\\
In Section II we revisit basic results around the concepts of \textit{instantaneous frequency} and \textit{group delay}, in particular a result related to the short-time Fourier transform (STFT). This revision shows the nature of the introduced concepts from a different perspective, which is the key for the main part of the paper. Analytical and numerical results are provided in order to strengthen the intuition. In Section III we start by briefly presenting the scattering transform in a time-frequency setting and adapt its principle structure to define the \emph{phase scattering coefficients}. Finally we show promising experiments on proof-of-concept examples in a direct comparison to (magnitude) scattering based on the STFT.
This first study sets the basis and intuition for further research on this idea.\\

The content of this paper was mainly developed in the Master's thesis of the first author \cite{mastersthesis}. More details, including proofs and \textsc{Matlab} scripts are available under  \url{https://github.com/danedane-haider/Phase-Scattering-Masterthesis}.
The scripts use the LTFAT toolbox \cite{ltfat,ltfatnote030}.

\section{Instantaneous Frequency and Group Delay}\label{sec:ifandgd}
\textit{Instantaneous frequency} (IF) and \textit{group delay} (GD) are important concepts in signal analysis \cite{flandrin} as they provide precise information about frequency, resp. temporal positioning of signal components \cite{ma09}.
%
%
%
Local versions are defined by the partial derivatives of the phase in the time-frequency plane. Let $\Phi_{\Psi}^x(t,\omega)$ denote the phase function of a complex-valued time-frequency representation, based on a family of localization functions $\Psi$, then
\begin{equation}\label{cif}
\widehat{\omega}_{\scriptscriptstyle \Psi}(x;t,\omega)=\frac{1}{2\pi}\frac{\partial}{\partial t}\Phi_{\Psi}^x(t,\omega)
\end{equation}
is called \emph{channelized instantaneous frequency} (CIF) and
\begin{equation}\label{lgd}
\widehat{\tau}_{\scriptscriptstyle \Psi}(x;t,\omega)=-\frac{\partial}{\partial \omega}\Phi_{\Psi}^x(t,\omega)
\end{equation}
\emph{local group-delay} (LGD) of $x$ w.r.t. $\Psi$. These concepts have been used already many times, e.g. for the phase vocoder \cite{phasevocoder} or reassignment methods \cite{tfreass,
holighprus16} and other types of phase-based processing \cite{pbproc}.
In this work we focus on the STFT as underlying time-frequency representation. According to the scaling conventions we use, its phase function $\Phi_{\scriptscriptstyle g}^x(t,\omega)$ using a window $g$ is found in its magnitude-phase representation in the following way,
\begin{equation}
    \mathcal{V}_gx(t,\omega) = |\mathcal{V}_gx(t,\omega)| \cdot e^{2\pi i \Phi_{\scriptscriptstyle g}^x(t,\omega)}.
\end{equation}
Usually, when working with the STFT, a specific phase convention is assumed. The resulting phase can then be converted into a time-invariant and a frequency-invariant version. In contrast to previous work, we select specific invariances in the STFT beforehand for computing CIF and LGD respectively. For $x,g\in \textbf{L}^2(\mathbb{R})$ the \emph{frequency-invariant} STFT is defined by
\begin{equation}\label{stftfreq}
\mathcal{V}_{g}x(t,\omega)=\int_{\mathbb{R}}x(\tau)\overline{g(\tau-t)}e^{-2\pi i\omega \tau}d\tau,
\end{equation}
$t,\omega \in \mathbb{R}$. Here, the exponential term is independent of the time index $t$. The \emph{time-invariant} STFT is defined by changing the order of modulation and translation, giving
\begin{equation}\label{stfttime}
\mathcal{V}_g^tx(t,\omega) = \int_{\mathbb{R}}x(\tau)\overline{g(\tau-t)}e^{-2\pi i\omega (\tau-t)}d\tau.
\end{equation}
Our approach is to consider CIF in a frequency-invariant \eqref{stftfreq} and LGD in a time-invariant setting \eqref{stfttime}. This results directly in modes of these quantities that are necessary for the last chapter of the paper. Note that this setting only agrees with the original definitions \eqref{cif} and \eqref{lgd} up to the induced invariances. Therefore, we will use the notation CIF\textsubscript{f} and LGD\textsubscript{t} further on.
The following proposition shows how CIF\textsubscript{f} and LGD\textsubscript{t} can be computed based on a result by Auger and Flandrin \cite{tfreass}. We additionally emphasize the formal requirements \cite{pole} and consider the introduced STFT-invariances \eqref{stftfreq} and \eqref{stfttime}.

\begin{proposition}\label{ciflgdstft}
Let $x \in L^2(\mathbb{R})$ and $g \in L^2(\mathbb{R}) \cap \mathcal{C}^1(\mathbb{R})$, such that $Tg \in L^2(\mathbb{R})$, where the time-weighted window $Tg$ is given by $Tg(\tau)=\tau g(\tau)$ for all $\tau\in\mathbb{R}$. Then $\mathcal{V}_gx(t,\omega)$ and $\mathcal{V}_g^tx(t,\omega)$ are infinitely partially differentiable in both variables, $t$ and $\omega$.
Let further $g'$ denote the differentiated window $g'(\tau)=\frac{d}{d\tau}g(\tau)$. Then, for $\mathcal{V}_gx(t,\omega)\neq0$, CIF\textsubscript{f} and LGD\textsubscript{t} can be computed by
	\begin{align}
	\widehat{\omega}_{\scriptscriptstyle \text{STFT}}(x;t,\omega)&=-\frac{1}{2\pi}\operatorname{Im} \Biggl\{ \frac{\mathcal{V}_{g'}x(t,\omega)}{\mathcal{V}_gx(t,\omega)}\Biggr\}\label{cifrep}\\
	\widehat{\tau}_{\scriptscriptstyle \text{STFT}}(x;t,\omega)&=\operatorname{Re} \Biggl\{\frac{\mathcal{V}^t_{Tg}x(t,\omega)}{\mathcal{V}^t_gx(t,\omega)} \Biggr\},\ \ t,\omega \in \mathbb{R}. \label{lgdrep}
	\end{align}
\end{proposition}
\noindent
A compact proof of this can be found in \cite{mastersthesis}.
The impact of the STFT invariances gets clear in the following.

\subsection{Analytical Shapes of CIF and LGD - Revisited}
Based on Proposition \ref{ciflgdstft}, one finds the CIF\textsubscript{f} of stationary sinusoids and the LGD\textsubscript{t} of Dirac impulse to be affine linear functions. The results may be considered as known, however, they show an alternative point of view via the induced STFT-invariances which is crucial for the construction of the phase scattering coefficients in the last section. 
\subsubsection{CIF\textsubscript{f} Analytically}
The CIF\textsubscript{f} representation of a stationary sinusoid (based on a frequency-invariant STFT) has linear behaviour along the frequency axis, centered at the IF of the sinusoid.
\begin{lemma}\label{ciflemma}
Let $x(\tau)=e^{2\pi i \xi_0 \tau}$ be a complex sinusoid with frequency $\xi_0$ and $g$ a window function with $g,Tg\in L^2(\mathbb{R})\cap L^1(\mathbb{R})$ such that the Fourier transform of $g$ has no zeros and the Fourier transform of $Tg$ only a single one at $0$.
Then Equation \eqref{cifrep} reduces to
	\begin{equation}\label{affinecif}
	\widehat{\omega}_{\scriptscriptstyle \text{STFT}}(x;t,\omega) = \xi_0-\omega
	\end{equation}
$t,\omega \in \mathbb{R}$.
\end{lemma}
Note that using a time-invariant STFT results in $\widehat{\omega}_{\scriptscriptstyle \text{STFT}}(x;t,\omega) = \xi_0$, yielding the IF directly, whereas here a frequency shift function is computed, measuring the distance to $\xi_0$ linearly.

\subsubsection{LGD\textsubscript{t} Analytically}
The LGD\textsubscript{t} representation of a Dirac impulse (based on a time-invariant STFT) has linear behaviour along the time axis, centered at the impulse.
\begin{lemma}\label{lgdlemma}
Let $x(\tau)=\delta(\cdot - \tau_0)$ where $\delta$ denotes a Dirac impulse and $g\in L^2(\mathbb{R})\cap \mathcal{C}^1$ with $g'\in L^2(\mathbb{R})$ where $g$ has no zeros and $g'$ only a single one at $0$. Then Equation \eqref{lgdrep} reduces to
	\begin{equation}\label{affinelgd}
	\widehat{\tau}_{\scriptscriptstyle \text{STFT}}(x;t,\omega) = \tau_0-t
	\end{equation}
$t,\omega \in \mathbb{R}$.
\end{lemma}
Also here, using the time-invariant version of the STFT is key to obtain this affine linear function and is not obtained when using the frequency-invariant STFT.
The results of the Lemmata can be deduced directly from Proposition \ref{ciflgdstft}, see \cite{mastersthesis}.

Thus, by choosing the specific STFT versions in this discriminative way, CIF\textsubscript{f} and LGD\textsubscript{t} are linear functions in frequency resp. time direction. We briefly illustrate this also numerically to strengthen the intuition.
 
\subsection{Numerical Shapes of CIF and LGD - Revisited}

\begin{figure}[t]
    \vspace{-0.2cm}
    \hspace{-0.55cm}
    \includegraphics[scale=0.22]{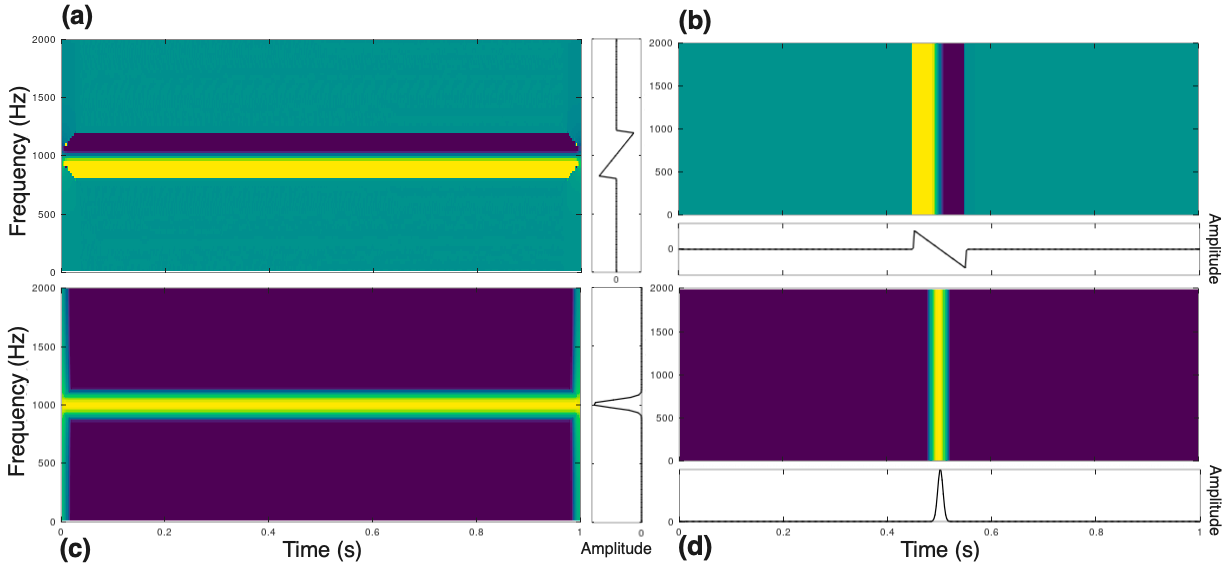}
	\caption{(a) Full CIF\textsubscript{f} of sinusoid and one column for fixed $t$ (right). (b) full LGD\textsubscript{t} of impulse and one column for fixed $\omega$ (bottom). (c) full STFT magnitude of sinusoid and one column for fixed $t$ (right). (d) full STFT magnitude of impulse and one column for fixed $\omega$ (bottom).}
	\label{fig:cif}
\end{figure}

Proposition \ref{ciflgdstft} provides a convenient method to compute phase derivatives numerically by point-wise operations of two STFTs.
However, when working with finitely supported windows in applications, the computations \eqref{cifrep} and \eqref{lgdrep} clearly have to be restricted to the support of the window, i.e. CIF\textsubscript{f} and LGD\textsubscript{t} coefficients are considered to be zero outside the supports. Therefore, in an actual numerical representation the linear shapes from \eqref{affinecif} and \eqref{affinelgd} are localized naturally to the range of the window support as well.
Using the toy examples from Lemma \ref{ciflemma} and \ref{lgdlemma} with $\xi_0=1000$Hz and $\tau_0=0.5$s, one finds the piece-wise linearly decreasing functions locally around $\xi_0$, resp. $\tau_0$.
Figure \ref{fig:cif} (a),(b) illustrates well how CIF\textsubscript{f} and LGD\textsubscript{t} resemble one another in terms of their shape in time, resp. frequency direction, emphasizing the duality of the two quantities. Panels (c),(d) show the direct comparisons to STFT magnitudes. Besides the intuition this brings for phase-based signal representations in general, the chosen STFT conventions allow to use CIF\textsubscript{f} and LGD\textsubscript{t} in a scattering procedure in a meaningful way.

For the calculation of the plots we used the 
function \texttt{gabphasederiv} from the \textit{Large Time-Frequency Toolbox} (LTFAT) \cite{ltfat} with the additional flags \texttt{'freqinv'} and \texttt{'timeinv'}, which follow the STFT conventions used in Proposition \ref{ciflgdstft}, based on the discrete Gabor transform \texttt{dgt}.
Discrete versions of a dilated Gaussian are used as window functions.

\section{Phase Scattering}

Indeed, the partial derivatives of the time-frequency domain phase are useful quantities for describing time-frequency information since they provide precise time and frequency localization of harmonic resp. transient events of a signal.
Using the analytical results and the intuition from the previous section, we aim for expanding the representations to higher orders in the manner of the \emph{scattering transform}.

\subsection{Time-Frequency Scattering}
The computational scheme of the scattering transform is a cascade of absolute values of filter transformations
\cite{scat,semi}.
A network-like structure emerges that results in layers of 
magnitude filter decompositions.
In a time-frequency setting this can be seen as taking one set of localization functions (band-pass filters) per layer, applied via convolution (originally using wavelet filters). Several papers show this transform as promising tool to represent large-scale features of audio signals, such as amplitude or frequency modulation, chord structures (intervals), tempo and rhythm, etc. \cite{deepscat,gabscat,paper}. Therefore, it is often used as external feature extraction step for deep learning tasks and moreover, is celebrated as being an interpretable neural network itself with fixed filters and a solid mathematical foundation, allowing for rigorous analysis. 
\begin{definition}[Time-Frequency Scattering]
    Let $x\in L^2(\mathbb{R})$ and $\Psi_{k}=\{\psi_i^{k}\}_{i\in\Lambda_{k}}$ denote the set of filters for the $k$-th layer. $\Lambda_{k}$ is the corresponding (frequency-)index set and the operator $\mathcal{U}_{k}[p_{0}]=|\psi_{p{0}}\ast \cdot\ |$ computes the magnitude filtering at index $p_{0}$.
    Then, the scattering transform is a cascade of the operators $\mathcal{U}_1,\dots ,\mathcal{U}_k$ along a frequency-index path $p=(p_1,\dots,p_{k})\in \Lambda_{1}\times \dots \times \Lambda_{k}$ and we call
    \begin{equation}
        \mathcal{S}[p]x = \mathcal{U}_{k}[p_{k}] \dots \mathcal{U}_{2}[p_{2}] \mathcal{U}_{1}[p_{1}]x
    \end{equation}
    the $k$-th order (magnitude) scattering coefficients w.r.t. $p$.
\end{definition}
The original idea of the scattering transform was to create a translation-invariant representation on $L^2(\mathbb{R})$ by integrating over the emerging scattering coefficients \cite{scat}. In applications, a final low-pass filtering step is used to simulate this property locally. In this work we preliminarily do not consider this step and focus merely on the nature of the coefficients.

\begin{figure*}[ht]
    \hspace{-0.55cm}
    \includegraphics[scale=0.228]{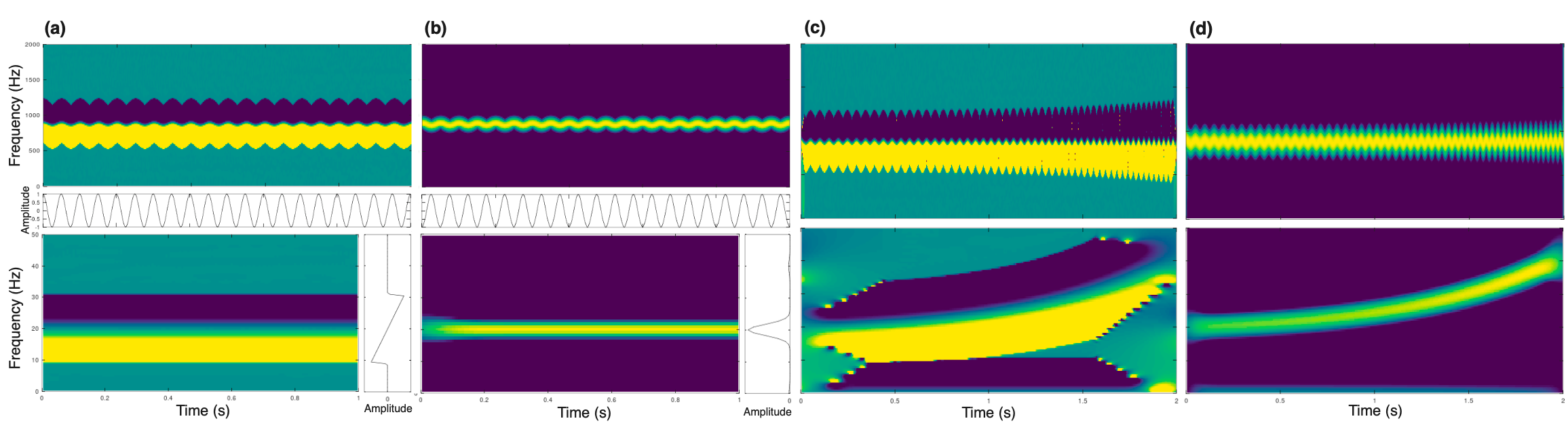}
	\caption{(a),(b) $1$st and $2$nd order CIF\textsubscript{f} and STFT scattering layers of $x_1$ with propagation channel at $880$Hz. (c),(d) $1$st and $2$nd order CIF\textsubscript{f} and STFT scattering layers of $x_2$.}
	\label{fig:cifcif}
\end{figure*}

\subsection{CIF and LGD Scattering}
Instead of taking the modulus of the complex time-frequency coefficients, we use the mappings to the partial derivatives of the phase as non-linearity in the scattering cascade. This procedure can be written in terms of operators, defined in the following way.
\begin{definition}[Phase Scattering Coefficients]
	Let $x\in L^2(\mathbb{R})$ and $\Psi$ a family of localization functions with index set $\Lambda$, then the operators
	\begin{align}
	\big(\widehat{\Omega}_{\scriptscriptstyle \Psi}[p_0]x\big)(t):=\widehat{\omega}_{\scriptscriptstyle \Psi}(x;t,p_0)\label{Omega}\\
	\big(\widehat{T}_{\scriptscriptstyle \Psi}[p_0]x\big)(t):=\widehat{\tau}_{\scriptscriptstyle \Psi}(x;t,p_0)\label{T}
	\end{align}
	compute CIF and LGD of $x$ at the frequency index $p_0\in\Lambda$ as function of time. Let $p=(p_1,\dots,p_{k})\in\Lambda_1\times \dots \times \Lambda_k$ denote a frequency-index path associated to $k$ families of localization functions $\Psi_1,\dots ,\Psi_k$, then cascading the operators \eqref{Omega} and \eqref{T} along $p$ defines the $k$-th order CIF scattering coefficients
	\begin{align}\label{cifscat}	
	\widehat{\Omega}[p]x=\widehat{\Omega}_{\scriptscriptstyle \Psi_k}[p_{k}]\dots\widehat{\Omega}_{\scriptscriptstyle \Psi_1}[p_1]x
	\end{align}
	and the $k$-th order LGD scattering coefficients
	\begin{align}\label{lgdscat}	
	\widehat{T}[p]x=\widehat{T}_{\scriptscriptstyle \Psi_k}[p_{k}]\dots\widehat{T}_{\scriptscriptstyle \Psi_1}[p_1]x.
	\end{align}
    of $x$ w.r.t. $p$. We shall call $\widehat{\Omega}[(q_1,...,q_{k-1},\lambda)]x$, considered for all $\lambda\in\Lambda_k$, the $k$-th CIF scattering layer of $x$ w.r.t. the path $q = (q_1,...,q_{k-1})$.
    Analog for the LGD case.
\end{definition}

Building on the results from the previous chapter, we use a phase scattering procedure that use CIF\textsubscript{f} and LGD\textsubscript{f} computed by the STFT to provide numerical examples. 

\begin{definition}[STFT-Phase Scattering Coefficients]
    Under the conditions from Proposition \ref{ciflgdstft}
	we can define the STFT-phase (or CIF\textsubscript{f} resp. LGD\textsubscript{f}) scattering coefficients analogously to \eqref{cifscat} and \eqref{lgdscat} by cascading the operators
	\begin{align}	
	\big(\widehat{\Omega}_{\scriptscriptstyle \text{STFT}}[p_0]x\big)(t)=
	-\frac{1}{2\pi}\operatorname{Im}\Biggl\{ \frac{\mathcal{V}_{g'}x(t,p_0)}{\mathcal{V}_{g}x(t,p_0)}\Biggr\},\label{cifauger}\\
	\big(\widehat{T}_{\scriptscriptstyle \text{STFT}}[p_0]x\big)(t)=
	\operatorname{Re}\Biggl\{\frac{\mathcal{V}^t_{Tg}x(t,p_0)}{\mathcal{V}^t_{g}x(t,p_0)} \Biggr\},\quad t\in \mathbb{R}\label{lgdauger}
	\end{align}
	along a frequency-index path $p=(p_1,...,p_{k})\in \mathbb{R}^k$, associated to $k$ STFTs using windows $g_1,\dots ,g_k$.
\end{definition}
Inspired by the experiments from \cite{gabscat,deepscat,paper}, we show that $2$nd order phase scattering coefficients are capable of capturing wider scale information, comparable to magnitude scattering.

\subsubsection{CIF\textsubscript{f} Scattering Application - Frequency Modulation}
We first show that CIF\textsubscript{f} scattering of a frequency modulated sinusoid finds the modulation frequency in the $2$nd layer.
\begin{lemma}\label{mod}
Let $x(\tau)=e^{2\pi i(\xi_0 \tau + \gamma(\tau))}$ with $\gamma(\tau) =\sin(2\pi \xi_1 \tau)$ denote a vibrato signal with center frequency $\xi_0$ and modulation frequency $\xi_1$. Then the $2$nd order CIF\textsubscript{f} scattering layer of $x$ w.r.t. $p_1=\xi_0$ is given by
\begin{equation}
	(\widehat{\Omega}_{\scriptscriptstyle \text{STFT}}[\xi_0,\omega ]x)(t)= \xi_1-\omega
\end{equation}
$t,\omega \in \mathbb{R}$.
\end{lemma}
This result shows the natural extension of the ($1$st order) CIF\textsubscript{f} representation from Lemma \ref{ciflemma}. Let $\phi(\tau)$ denote the temporal phase of $x$, given as above. We see that the instantaneous frequency of $x$ is  $\omega(x;\tau)=\frac{1}{2\pi}\frac{d}{d \tau}\phi(\tau)= \xi_0 + 2\pi \xi_1\cos(2\pi \xi_1\tau)$, which when considered as a time-domain signal itself possesses the IF $\omega(\omega(x;\tau);\tau) = \xi_1$, the modulation frequency of the vibrato.
Using $2$nd order CIF\textsubscript{f} scattering, we find $\xi_1$ as zero of an affine linear function, coming from $\big(\widehat{\Omega}_{\scriptscriptstyle \text{STFT}}[\omega]x\big)(t) = \omega(x;t) - \omega$. A more detailed discussion can be found in \cite{mastersthesis}.

Note that this result theoretically does not depend on the chosen propagation index $p_1$, however, in the numerical case we have to care about staying inside the support of the window, i.e. $p_1 \in \operatorname{supp}(\widehat{g})$.

As numerical examples we set up two sinusoidal signals $x_1,x_2$ at $880$Hz and modulate $x_1$ by $\gamma_1(\tau) = \sin(2\pi 20 \tau)$, creating a constant vibrato of $20$Hz and $x_2$ by $\gamma_2(\tau) = \sin(2\pi \tau(20+e^{\tau}))$, creating a vibrato that increases exponentially in frequency, beginning at $20$Hz. Figure \ref{fig:cifcif} shows $1$st and $2$nd order CIF\textsubscript{f} and STFT scattering layers of $x_1$ and $x_2$, using the propagation frequency index corresponding to $880$Hz. Although the modulation frequencies induced by $\gamma_1$ and $\gamma_2$ are extremely low, they are captured precisely in the $2$nd CIF\textsubscript{f} layers respectively as zeros of linear functions. This can be observed also for more general frequency modulation functions. Magnitude scattering finds the modulation frequencies as well, but here in terms of shifted window functions.
\subsection{Mixed Phase Scattering}
To provide an intuitive application of LGD scattering, we refer to an earlier work \cite{paper} where we have shown how STFT magnitude scattering can capture the frequency of temporal patterns among transients in a signal. In this sense, our approach is to use the LGD\textsubscript{t} representation as transient detector and use a subsequent CIF\textsubscript{f} transformation to get the frequency information of the LGD\textsubscript{t}-peak arrangement.

\begin{definition}[Mixed Phase Scattering]
	Let $x\in L^2(\mathbb{R})$ and $\Psi_1, \Psi_2$ families of localization functions, then we define
	\begin{equation}
	\widehat{M}[p_1,p_2]x= \widehat{\Omega}_{\scriptscriptstyle \Psi_2}[p_2]\widehat{T}_{\scriptscriptstyle \Psi_1}[p_1 ]x
	\end{equation}
	as $2$nd order mixed phase scattering coefficients of $x$ w.r.t. the path $(p_1,p_2)$.
\end{definition}

\subsubsection{Mixed Phase Scattering Application - Dirac Comb}
\begin{figure}[t]
    \vspace{-0.22cm}
    \hspace{-0.55cm}
    \includegraphics[scale=0.252]{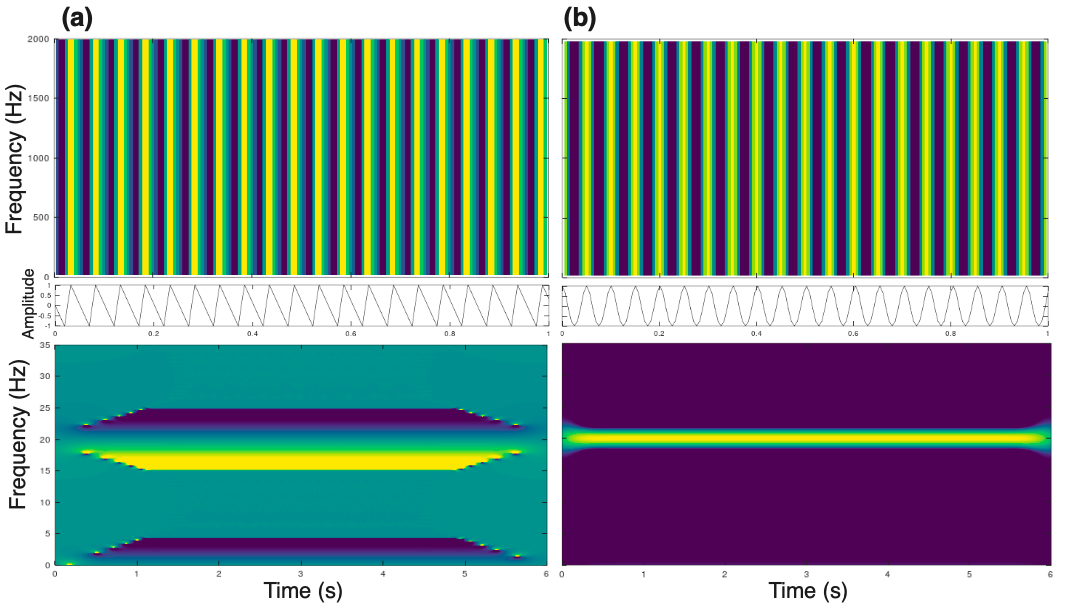}
	\caption{(a) $1$st and $2$nd order mixed phase scattering layers of the Dirac comb w.r.t. an arbitrary $p_1$. (b) $1$st and $2$nd order STFT scattering layers.}
	\label{fig:mix}
\end{figure}

We show that $2$nd order mixed scattering finds the fundamental frequency of a Dirac comb signal.
\begin{lemma}
Let $x(\tau)=
	\sum_{\ell\in\mathbb{Z}}\delta(\cdot - \frac{2 \pi \ell}{\xi_0})$ be a Dirac comb with fundamental frequency $\xi_0$. Then the $2$nd order mixed phase scattering layer of $x$ is given by
\begin{equation}
   \big(\widehat{M}[p_1,\omega]x\big)(t)=\xi_0-\omega
\end{equation}
for any $p_1$ and $t,\omega \in \mathbb{R}$.
\end{lemma}
Crucial for this result is to use a window $g$ with $\operatorname{supp}(g) \geq 2\pi/\xi_0$ to avoid regions of zero amplitude between the impulses. Then $\big(\widehat{T}_{\scriptscriptstyle \text{STFT}}[\omega]x\big)(t)$ is a saw-tooth wave for every $\omega\in\mathbb{R}$ with a fundamental frequency of $\xi_0$ and the same phase function as its sinusoidal equivalent, i.e. the IF is found by the subsequent CIF\textsubscript{f} transformation analog to Lemma \ref{mod}. A more detailed discussion can be found in \cite{mastersthesis}. Figure \ref{fig:mix} shows the mixed phase scattering and STFT scattering layers of a Dirac comb signal with fundamental frequency of $20$Hz.

\section{Conclusion and Outlook}
We revisited channelized instantaneous frequency and local group delay by pointing out their linear characters when considering two different STFT-invariance settings. Analytical and numerical arguments are provided in order to emphasize this perspective. Based on that we introduced an extension of these representations in the manner of the scattering transform by cascading the computations, called \emph{phase scattering}. It turned out that this procedure is, like the original scattering, capable of extracting information that lives in larger scales. We found modulation frequencies of vibrato signals and the tempo of a Dirac comb emerging in the second layer of the transform. However, other than the blurred representation of time-frequency information that magnitude scattering provides, CIF\textsubscript{f} and LGD\textsubscript{t} scattering allows for precise allocation of large-scale time-frequency information by means of zeros of linear functions.

This first study provides the foundation and intuition for further investigations on scattering phase information. The many existing results related to the scattering transform may serve as source of inspiration. In particular, to provide a mathematically well-defined setting one could formulate phase scattering in terms of distribution theory. Then, the expected properties regarding translation-invariance could be approached in theory. An immediate next step for continuing numerical experiments should be the development of computational algorithms that produce clean plots and further on, how phase scattering performs on real world audio signals. Finally, given the precise allocation properties, it seems promising to evaluate a set of phase-based scattering features as pre-processing step for a classification task.

\section*{Acknowledgment}
The work on this paper was partially supported by the Austrian Science Fund (FWF) START-project FLAME (Frames and Linear Operators for Acoustical Modeling and Parameter Estimation; Y 551-N13) and MERLIN (I 3067–N30).

\bibliographystyle{IEEEbib}

\bibliography{references,bibliopeter}

\end{document}